\documentclass{IEEEtran}

\usepackage[normalem]{ulem}
\usepackage[hyphens]{url}
\usepackage{graphicx}
\usepackage[bookmarks=false]{hyperref}
\usepackage{subcaption}
\usepackage{algorithm, algorithmic}
\usepackage{blindtext}
\usepackage{cite}
\usepackage{color}   
\usepackage{amssymb,amsmath}
\usepackage{comment} 
\usepackage{enumerate}
\usepackage{framed}
\usepackage{array}      
\usepackage{url}		
\usepackage{balance}
\usepackage{xspace}
\usepackage{url}         
\usepackage{multirow}
\usepackage{listings}
\usepackage{booktabs}
\usepackage{amsmath}
\usepackage{datetime}
\usepackage{balance}
\newcommand{\subparagraph}{}
\usepackage[compact]{titlesec}

\hypersetup{bookmarks={false}}

\newcommand{\code}[1]{\texttt{#1}}

\title{Similarity Search on Automata Processors} 
\author{Vincent T. Lee, Justin Kotalik, Carlo C. del Mundo, Armin Alaghi, Luis Ceze, Mark Oskin\\
University of Washington\\
\{vlee2, jkotalik, cdel, armin, luisceze, oskin\}@cs.washington.edu}

\begin{document}
\sloppy
\maketitle


\begin{abstract}

Similarity search is a critical primitive for a wide variety of applications including natural language processing, content-based search, machine learning, computer vision, databases, robotics, and recommendation systems.
At its core, similarity search is implemented using the k-nearest neighbors (kNN) algorithm, where computation consists of highly parallel distance calculations and a global top-k sort.
In contemporary von-Neumann architectures, kNN is bottlenecked by data movement which limits throughput and latency.
In this paper, we present and evaluate a novel automata-based algorithm for kNN on the Micron Automata Processor (AP), which is a non-von Neumann near-data processing architecture.
By employing near-data processing, the AP minimizes the data movement bottleneck and is able to achieve better performance.
Unlike prior work in the automata processing space, our work combines temporal encodings with automata design to augment the space of applications for the AP.
We evaluate our design's performance on the AP and compare to state-of-the-art CPU, GPU, and FPGA implementations; we show that the current generation of AP hardware can achieve over 50$\times$ speedup over CPUs while maintaining competitive energy efficiency gains.
We also propose several automata optimization techniques and simple architectural extensions that highlight the potential of the AP hardware.
\end{abstract}


\section{Introduction}

The growth of media content combined with advances in machine learning, vision, and robotics has catalyzed the growth and demand for applications such as document retrieval~\cite{document_distances}, content-based search~\cite{content_based_search}, and deduplication~\cite{dedup}.
In 2010, Internet users had shared more than 260 billion images on Facebook~\cite{facebook-photos}, and, by 2014, uploaded upwards of 300 hours of YouTube per minute~\cite{youtube-statistics}.
Furthermore, the rate of multimedia content generation is projected to grow exponentially~\cite{rebooting-the-it-revolution}.

To make these volumes of data searchable, applications rely on similarity search which manifests as k-nearest neighbors (kNN)~\cite{google-images, facebook-search}.
The kNN algorithm consists of many parallelizable distance calculations and a single global top-k sort.
While computationally very simple, scaling kNN beyond a single node is challenging as the algorithm is memory bound in both CPUs and GPUs.
Distance calculations are relatively cheap and task parallel but moving feature vector data from memory to the compute device is a huge bottleneck.  
Moreover, this data is used only once per kNN query and discarded, and the result of a kNN query is only a handful of identifiers. 
In order to amortize data movement cost, queries are typically batch processed which has intrinsic limitations when constrained to meet tight latency targets. 

Because of its significance, generality, parallelism, simplicity, and small result set, kNN is an ideal candidate for near-data processing.
The key insight is that by exploiting higher bandwidths and applying large data reductions near memory we can substantially reduce data movement requirements over interconnects such as PCIe.
There have been many proposals for near-data processing and processing-in-memory (PIM) in the past~\cite{McKee95,iram,flexram,activepages}. 
In this paper, we evaluate kNN on the Micron Automata Processor (AP), a recent PIM architecture for high speed automata evaluation.
Most prior work on the AP has focused on the domain of pattern mining such as biological motif search~\cite{ap-motif-search}, frequent itemset mining~\cite{ap-apriori}, and graph processing~\cite{ap-graph}.

In contrast, our work is the first to explore and evaluate similarity search on the AP which presents unique design challenges and opportunities.
We propose a novel nondeterministic finite automata design for kNN using temporal encodings and show current generation hardware can achieve $\sim$50$\times$ performance over multicore processors.
We also evaluate AP performance and energy efficiency against competing FPGA and GPU solutions.
We then propose mutually orthogonal optimization techniques and architectural extensions which can yield an additional $\sim$70$\times$ potential performance improvement which can make the AP a competitive alternative to existing heterogeneous computing substrates.

Our work makes the following contributions: (1) We present a novel automata design which exploits temporal encodings to solve the sorting phase of kNN in linear time with respect to dimensionality (as opposed to linear time with respect to dimensionality \emph{times} cardinality). (2) We introduce automata design optimizations that can reduce the resource footprint and improve throughput on future generation APs. (3) We propose and evaluate the potential of architectural extensions to augment the capabilities and efficiency of the AP architecture.

The paper is organized as follows. 
Section~\ref{sec:background} provides background on the AP and kNN. 
Section~\ref{sec:automata-design} presents our novel kNN automata design and Section~\ref{sec:methodology} outlines evaluation methodology.
Section~\ref{sec:evaluation} presents performance and energy efficiency results.
Section~\ref{sec:optimizations} presents automata optimization techniques, Section~\ref{sec:analysis} proposes AP architectural extensions, and Section~\ref{sec:related-work} highlights related work.

\section{Background} \label{sec:background}

\subsection{k-nearest neighbors} 
\label{sec:knn}

Performing similarity search between a query and a dataset of candidate vectors is done through k-nearest neighbors (kNN), and consists of distance calculations and global top-k sorting (Algorithm~\ref{knn_algorithm}).
Intuitively, the goal of kNN is to find the closest neighbors for a \textit{query vector} against a set of candidate vectors in a database. 
A query and candidate vector are considered more similar if the metric space distance between them is smaller.
For instance, two images in content-based search~\cite{content_based_search} are considered more similar if the distance between their feature descriptor vectors is smaller.
Feature vectors are generated by applying feature extractors like word embeddings~\cite{document-distances}, scale invariant feature transforms (SIFT)~\cite{sift}, or semantic embeddings~\cite{tagspace} to target media.
The exact kNN algorithm scans through the dataset for each query/candidate pair resulting in a total complexity of $O(qnd)$ where $q$ is the number of query vectors, $n$ is the number of dataset vectors, and $d$ is the vector dimensionality.
The algorithm presents two opportunities to exploit parallelism: query level parallelism where multiple queries are batch processed, and data level parallelism where multiple dataset vectors are processed in parallel for a single query.

\begin{algorithm}
\begin{algorithmic}
    \FOR {each query $q \in Q$}
       \STATE Result set $R \leftarrow \emptyset$
       \FOR {each point $p \in$ dataset $D$}
          \STATE $d \leftarrow distance(p, q)$
          \STATE $R = R \cup (ID = p, dist = d)$
       \ENDFOR
       \STATE Sort $R$ by $dist$
    \RETURN Top $k$ points $\in R$ with smallest $dist$
    \ENDFOR
\end{algorithmic}
\caption{kNN}
\label{knn_algorithm}
\end{algorithm}

Approximate variants of kNN employ spatial indexing structures to trade accuracy for run time by using heuristics to prune the search space.
We consider three approximate kNN algorithms: randomized kd-trees, hierarchical k-means, and locality sensitive hashing~\cite{flann, lsh}.
In randomized kd-trees, the dataset is indexed across multiple parallel trees which each partition the dataset based on dimensions with highest variance.
The index structure size scales exponentially with depth so the height of the tree is often constrained.
Each leaf in the tree is associated with a bucket of candidate points; when a traversal reaches a leaf, the associated bucket is linearly scanned to compute the result.
Similarly, hierarchical k-means generates an index by hierarchically partitioning the dataset into clusters.
Unlike randomized kd-trees, traversing the k-means index requires a distance calculation at each node to determine the next traversal.
Again, each leaf in the index represents a bucket of candidate points which is scanned linearly after the traversal.
Finally, locality sensitive hashing (LSH) uses a set of hash functions to hash similar query vectors into the same bucket~\cite{lsh}; the bucket each vector hashes to is again linearly scanned at search time.

Automata processors do not have hardened arithmetic units so implementing Euclidean distance is inefficient.
Instead, we subset our design space by evaluating Hamming distances which map well to the AP.
Real-valued vectors can be converted to Hamming space by applying techniques like iterative quantization (ITQ)~\cite{gong2011ITQ}. 
Some information in the features is lost as quantization narrows the possible dynamic range of feature values.
In practice, Hamming codes have been shown to be a viable alternative to Euclidean space encodings~\cite{lazebnik09, wang12, facebook-search}.
For instance, work by Lin et al.~\cite{lin16} shows that despite the loss of accuracy, well-crafted Hamming 
codes can achieve competitive or even better results compared to those of full precision representations.
In this paper, we assume dataset vectors are quantized offline using techniques like ITQ which removes this process from the critical path of the kNN kernel.

\subsection{Micron Automata Processor}
\label{sec:ap-description}

{
\setlength{\belowcaptionskip}{-15pt}
\begin{figure*}
\centering
\includegraphics[width=0.9\linewidth]{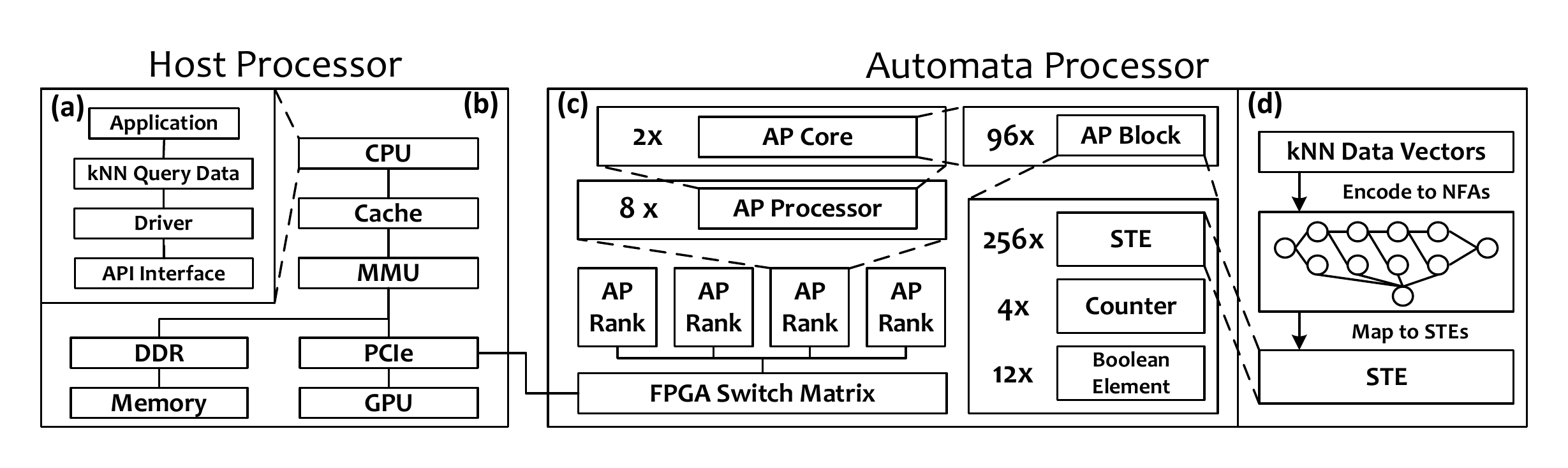}
\caption{System architecture of AP for kNN. (a) Software stack for an AP application. (b) Host processor system. (c) Internal architecture of AP device. (d) Data vectors are encoded into NFAs and mapped to STEs.}
\label{system-architecture}
\end{figure*}
}

\noindent\textbf{Hardware architecture.} The AP is a non-von Neumann architecture which uses a nondeterministic finite automata (NFA) driven execution model~\cite{MAP}; the system architecture is shown in Fig.~\ref{system-architecture}. 
A typical AP device is composed of four \textit{ranks} each containing eight automata processors. The eight automata processors are further subdivided into two half cores (AP cores).
Each half core is composed of 96 \textit{AP blocks} and each block contains 256 state transition elements (STEs); this results in a maximum of 24,576 STEs per half core or 1,572,864 STEs per device.
In the context of an NFA, each STE implements one state.
Since NFAs cannot span AP cores, the maximum size automata that can be implemented is limited to 24,576 states.

A central reconfigurable routing matrix (not shown) is responsible for connecting STEs to implement different NFAs.
Each AP block also contains four counters, 12 boolean elements, and a maximum of 32 reporting STEs.
Counters have an increment-by-one and reset port. The increment-by-one port increments the internal count when the connected state is active while the reset port resets the internal count.
Counters are always programmed to user-specified threshold values and activate downstream states if the internal count exceeds this static threshold; counters cannot be incremented by more than one and do not expose internal count values.
Finally, each AP block contains \textit{boolean elements} which can each be programmed to function as any standard two input logic gate.

\noindent\textbf{Programming model.} To program the AP, applications must be converted to equivalent NFAs expressed in terms of states, counters, and boolean elements.
Applications can either be compiled to NFAs by supplying a Perl Compatible Regular Expression (PCRE), or an XML-based Automata Network Markup Language (ANML) file which contains an NFA specification.
For applications that cannot be easily expressed as PCREs, the programmer must specify an ANML file.
The primary strength of the AP architecture is its capacity for high internal bandwidth and parallel computing; to fully harness this compute and bandwidth, it is ideal to instantiate many NFAs in parallel.

Each state in the NFA is associated with an 8-bit symbol or set of 8-bit symbols defined by a PCRE.
A state is activated when the input symbol matches the symbol set associated with it and any previous state connected to it is active on the previous time step.
The sequence of symbols is driven by a \textit{symbol stream} from the host processor; NFA state activations themselves cannot be combined to form a new symbol stream from inside the AP.
Each NFA also has specially designated \textit{start states} and \textit{reporting states} which are used to initiate state activations and return results from the AP respectively.
Start states do not need an upstream state to be active when matching the input symbol, and activate whenever the input symbol matches its symbol set.
Reporting states generate a signal which returns a unique identification number associated with it, and the offset (i.e. cycle accurate time stamp) within the symbol stream at which the state was activated.
This information is then used by a host processor application to resolve the result of the calculation.

\noindent\textbf{System integration.} An AP device interfaces to a host processor over a PCIe x8 Gen 3 interface (much like a GPU or FPGA).
The host processor is responsible for configuring the device, driving the symbol streams, and processing the signals generated by the reporting states.
Typically a host processing application will use provided APIs to operate the AP.

\subsection{Automata Design Constraints} 
\noindent\textbf{Arithmetic operations are inefficient.} The AP fabric does not contain hardened arithmetic units.
While STEs can be treated as lookup tables, implementing arithmetic units with lookup tables is inefficient and difficult to program.
This makes it difficult for the AP to calculate Euclidean distances.
On the other hand, Hamming distance and Jaccard similarity on the AP is well-documented and can be efficiently implemented~\cite{ap-hamming-distance}.

\noindent\textbf{No nested finite automata.} The programming model for the AP assumes that the symbol stream that drives NFAs is provided exclusively by an external host processor.
This means, state activations or output values from an NFA cannot be combined to dynamically form a symbol stream to drive another nested NFA.

\noindent\textbf{Symbol streams cannot be dynamically modified.} The external symbol streams that drive the NFAs in the AP cannot be modified; thus symbol streams must be defined statically before being streamed into the AP.
This eliminates any feedback interactions between the NFAs and symbol stream, making constructs such as dynamically inserting a reset symbol into the symbol stream impossible.

\section{Proposed Automata Design} \label{sec:automata-design}

We now present our automata design for Hamming distance kNN which is subdivided into two components: \textit{Hamming macros} for distance calculations and \textit{sorting macros} to sort distance scores.

\subsection{Hamming Macros}

The highly parallelizable distance calculation tasks in kNN are an excellent match for the AP.
In von-Neumann architectures, typically the dataset vectors are streamed from memory and the number of calculations performed is limited by the rate at which vectors can be streamed from memory; for $n$ vectors each with $d$ dimensions this takes $O(nd)$ time to complete all the distance calculations per query. 
For the AP we instead stream the queries to the dataset vectors encoded in NFAs on the AP, and perform comparisons against all candidate vectors in parallel much like a ternary content addressable memory.
This allows the AP to exploit the theoretical maximal degree of data parallelism in the context of kNN and reduces the run time to $O(d)$ per query; the only limitation is the number of spatial resources required to implement the large number of parallel automata.

\begin{figure}[!b]
\centering
\includegraphics[width=\linewidth]{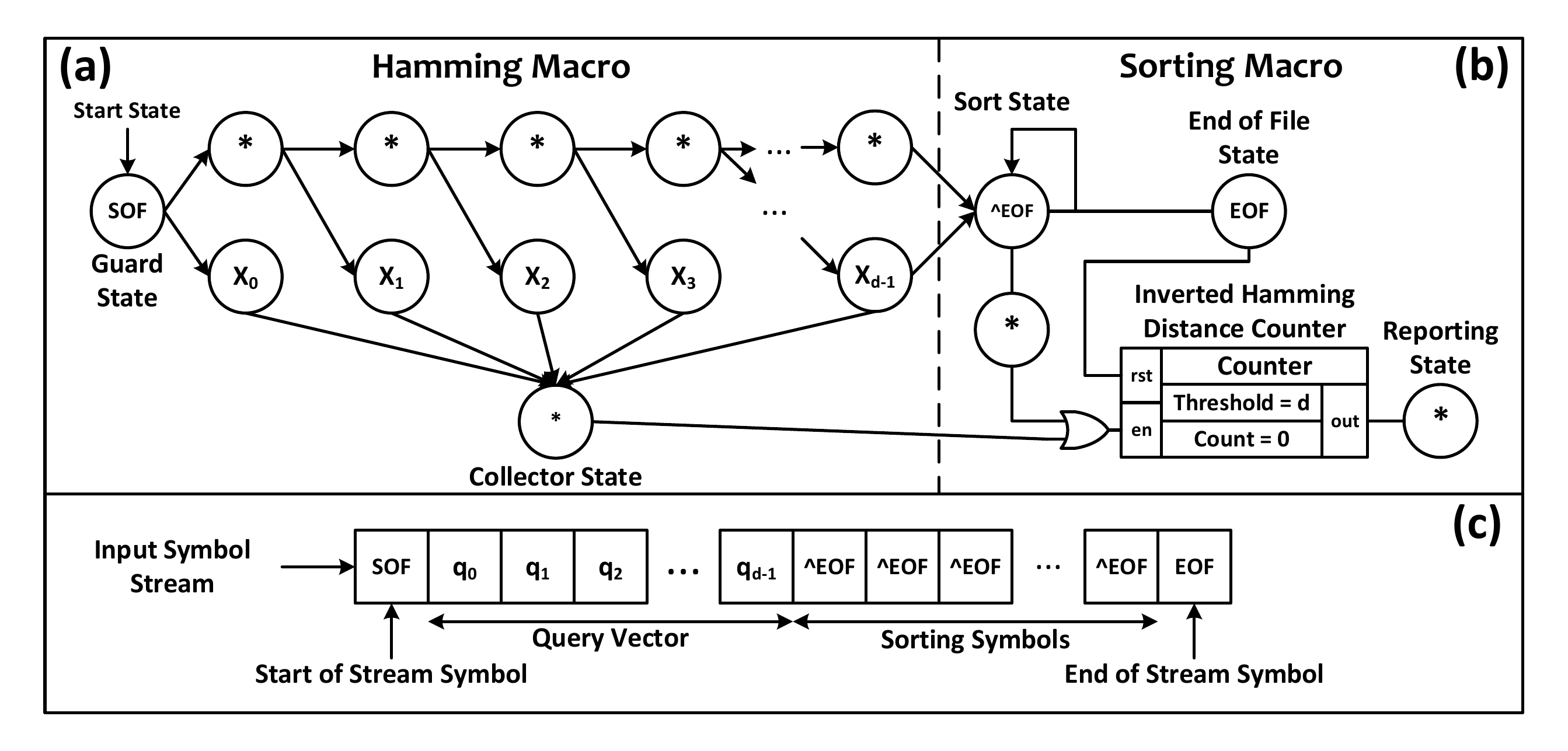}
\caption{Automata design encoding the vector $x = \{x_0, x_1, x_2, x_3, .., x_{d-1}\}$. (a) Hamming macro. (b) Sorting macro. (c) Symbol stream for a query $q = \{q_0, q_1, q_2, ..., q_{d-1}\}$.} 
\label{full-hamming-macro}
\end{figure}
 
We instantiate one Hamming macro per vector NFA in the dataset as shown in Fig.~\ref{full-hamming-macro}a; each Hamming macro computes the \textit{``inverted Hamming distance''} or the number of dimensions minus the Hamming distance. 
Effectively, this measures the similarity between the feature vector and the query.
Each Hamming macro encodes a unique feature vector from the dataset and operates in parallel with all other Hamming macros.
Since Hamming macros can be replicated arbitrarily, the factor of parallelism we can achieve is only limited by the capacity of the AP.
The Hamming macro is composed of a \textit{guard state}, \textit{compute states}, and \textit{collector states}.
The guard state is a designated start state which is responsible for detecting the start of file (SOF) symbol; this SOF symbol demarcates the start of an input query vector and protects the rest of the NFA from unintentional state activations. 
The compute states are composed of a sequence of ``*'' PCRE states\footnote{``*'' states match any symbol.} and matching states; each matching state activates when the input query vector value matches for the dimension in the encoded dataset vector.
To compute the inverted Hamming distance, states which correspond to matches are first fed into one or more collector states that effectively function as a large OR reduction.
The collector states then drive a counter which we refer to as a \textit{inverted Hamming distance counter} in the sorting macro. This counter keeps track of how many matches occurred.
For larger dimensional vectors we implement the collector states as a reduction tree of ``*'' states to limit the maximum state fan in and improve routability. 

\subsection{Sorting Macros}

The second phase of kNN consists of distilling the distance scores across every element in the dataset to the top $k$ neighbors.
Typically on von-Neumann architectures this is done using priority queue insertions which takes $O(n\log n)$ time per query, or alternative algorithms like k-selection.
For the AP, this must be conveyed using the activation signals from reporting states which consist of the unique state ID that map back to the feature vector, and the cycle offset into the symbol stream on which the state activated.
Furthermore, the sorting latency must be comparable to that of the Hamming macros otherwise it will dominate execution time.

We use a temporally encoded sort inspired by~\cite{race-logic} and~\cite{spaghetti-sort} which yields an $O(d)$ time sort.
Our sorting approach is based on the fact that vectors with lower Hamming distances have a higher inverted Hamming distance score.
To sort the vectors in parallel, we uniformly increment the inverted Hamming distance counters until they reach a static threshold equal to $d$; 
we use the pulse mode on the counter so that when the threshold is reached, a single cycle pulse activation is emitted from the counter.
Hamming macros which are most similar to the query vector have a higher inverted Hamming distance and activate their reporting states first.
As a result, the temporal order of the reporting state activations is sorted by increasing Hamming distance.
Note that the total number of cycles required to perform this temporally encoded sort is $O(d)$ making the sort latency the same as the Hamming macro latency.

To support temporal sort, a \textit{sorting macro} shown in Fig.~\ref{full-hamming-macro}b is appended to the Hamming macro.
To provide time for the sort to execute, we pad the symbol stream with filler symbols (\^{}EOF) for $O(d)$ cycles (Fig.~\ref{full-hamming-macro}c). 
A \textit{sort state} is responsible for waiting until an end of file (EOF) symbol is detected before triggering the EOF state which resets the counter.
The sort state also doubles as an unconditional increment signal for the inverted Hamming distance counter to perform the temporally encoded sort.
A \textit{reporting state} after the counter is responsible for reporting at which offset the inverted Hamming distance counter exceeded the threshold and the unique state ID assigned to that reporting state.
The unique state ID is used to reverse lookup which dataset vector it corresponds to before returning the result.

\begin{figure*}
\centering
\includegraphics[width=\linewidth]{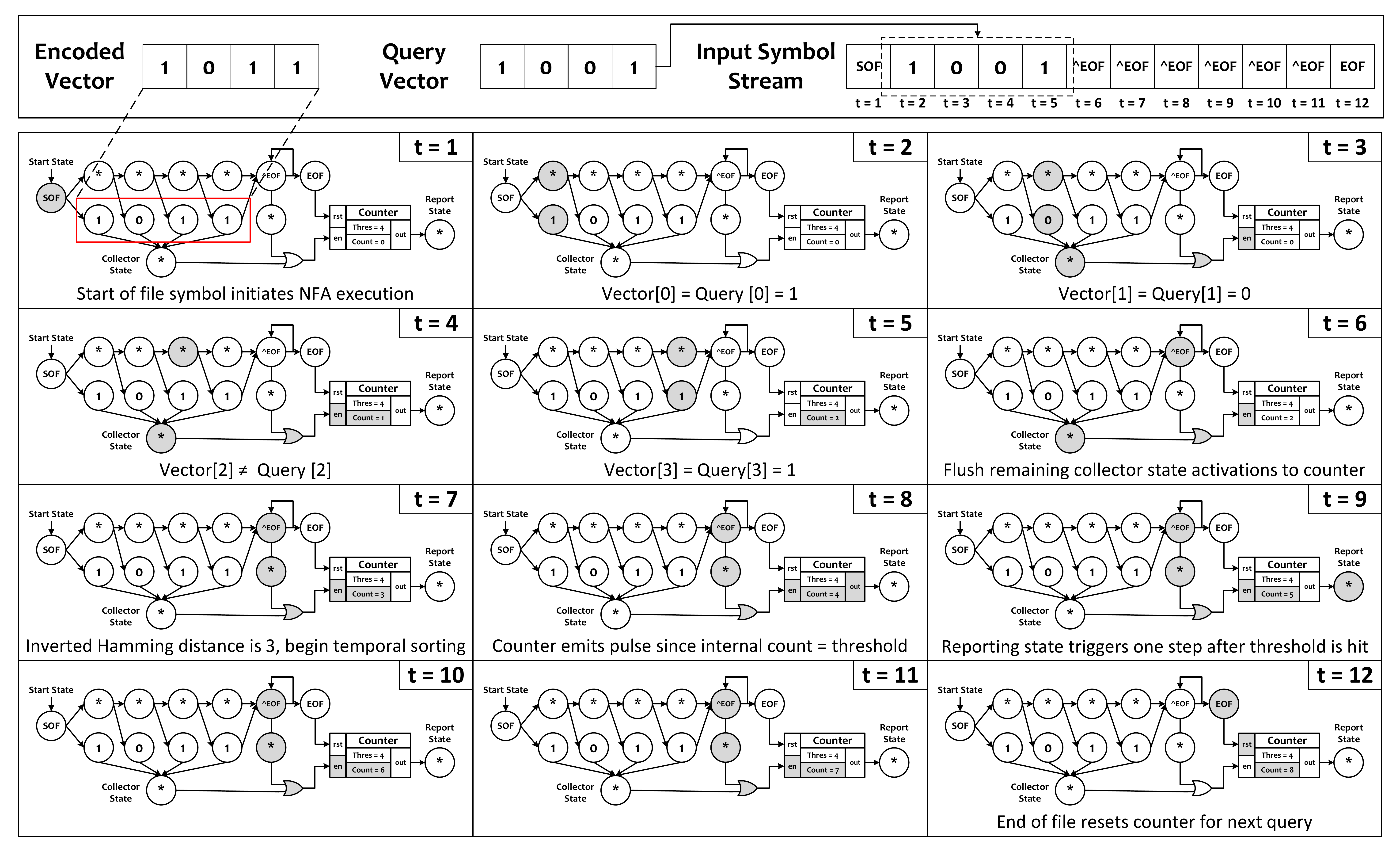}
\caption{Example execution for an NFA encoding the data vector \{1, 0, 1, 1\} and a symbol stream encoding the query vector \{1, 0, 0, 1\}. Grey states indicate active states or ports at each time step. The counter activates at time step t = 8 and emits a single activation pulse to the reporting state which activates the next cycle (t = 9).}
\label{large-example}
\end{figure*}

The concrete behavior of a \textit{single} combined Hamming and sorting macro is best illustrated with an example execution shown in Fig.~\ref{large-example}.
In this example, we show the state activations at each time step for an NFA which encodes the data vector \{1, 0, 1, 1\} and a symbol stream which encodes the query vector \{1, 0, 0, 1\}.
The activations during time steps 2 to 6 drive the Hamming distance calculation while time steps 7 to 11 drive the temporally encoded sort by uniformly incrementing the inverted Hamming distance counters towards the threshold.
The temporally encoded sort across vector NFAs is illustrated in Fig.~\ref{race-sort} which shows the NFA execution of two different data vectors.
In this example, vector $A$ has an inverted Hamming distance of 3 while vector $B$ has an inverted Hamming distance of 2.
During the sorting phase, the counter for vector $A$ reaches the target threshold $d$ first and triggers the reporting state for vector $A$ before vector $B$.
The resulting temporal order of the reporting state activations thus correctly encodes that $A$ has a lower Hamming distance than $B$ since $A$ reported first.

{
\setlength{\belowcaptionskip}{-10pt}

\begin{figure*}
\centering
\includegraphics[width=\linewidth]{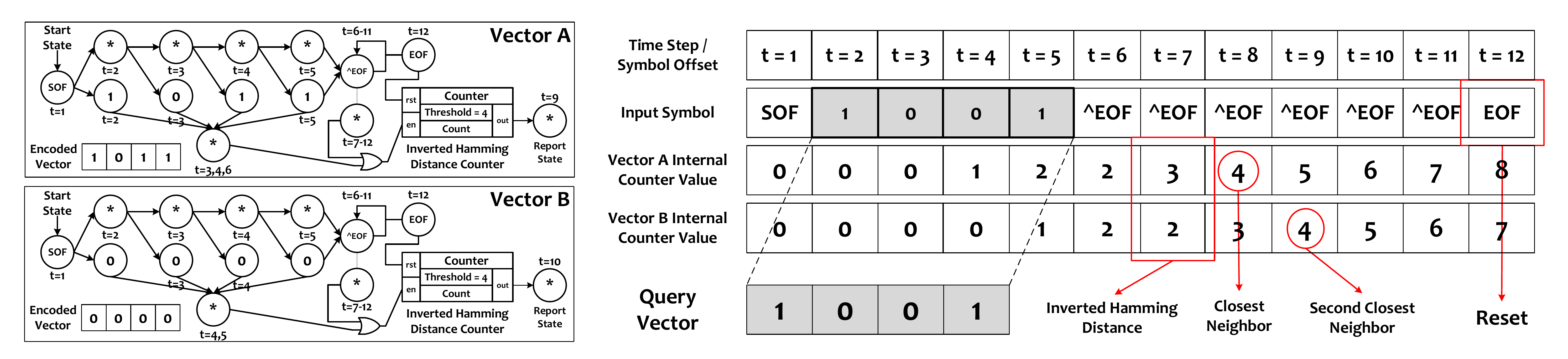}
\caption{Temporally encoded sort for two vectors A = \{1,0,1,1\} and B = \{0,0,0,0\} for a query vector C = \{1,0,0,1\}. The time step at which each state is activated is indicated next to each state. Vector A activates its reporting state first because it has a higher inverted Hamming distance.}
\label{race-sort}
\end{figure*}
}

\subsection{Partial Reconfiguration}
\label{sec:partial-reconfiguration}
The STE resources required by our design scales with both dataset size and vector dimensionality.
This means the capacity of the AP limits the total number of vectors that can be processed per configuration.
To scale to arbitrarily large datasets, we exploit the reconfiguration capability of the AP which allows the AP to be reconfigured much like an FPGA.
We assume these additional configurations are precompiled into a set of board images.
Once the queries are streamed through the first dataset partition, a reconfiguration prepares the AP with the next dataset partition until the entire dataset is processed.
This requires the host processor to also keep track of intermediary results per query across board reconfigurations as sort results are computed.
Current generation APs (Gen 1) require 45 ms~\cite{ap-apriori} per reconfiguration but next generation APs (Gen 2) are projected to be two orders of magnitude faster ($\sim$100$\times$)~\cite{ap-personal} which is closer to that of production-grade FPGAs~\cite{papadimitriou11}.

\subsection{Implementing Spatial Indexing Structures}

So far our automata design has been constrained to implementing linear kNN search.
While some index traversals are possible to express as automata, it is more efficient to factor the index traversal out to the host processor in software.
Since the AP is a spatial architecture, every encoded vector NFA needs to evaluate whether it is part of the pruned search space by traversing an index NFA.
In practice, only a few index traversals per query will be relevant making a vast majority of the traversals unnecessary. 
Instead, the host processor can traverse the index and pick which set of vector NFAs to load and query.
This modification is synergistic with kNN since the number of dataset vectors supported by each AP board configuration provides a natural bucket size limit for each indexing structure.
Finally, by offloading index traversal to the host processor, we can also support any arbitrary spatial indexing structures.

\section{Evaluation Methodology} \label{sec:methodology}

To evaluate the performance and energy efficiency gains of the AP, we compare against the CPUs, GPUs, and FPGAs shown in Table~\ref{evaluated-platforms}.
The Gen 1 AP board is not a production-ready platform so in order to provide a fair evaluation of the AP's potential, we provide a Gen 2 performance evaluation with better partial reconfiguration times. We also provide a more aggressive evaluation using optimizations and extensions proposed in later sections (AP Opt+Ext). 
AP Gen 2 more accurately reflects the capabilities of a production ready platform while AP Opt+Ext reflects an upper bounds with design optimizations and architectural extensions.

\begin{table}
\centering
\caption{Evaluated platforms.}
\begin{tabular}{@{}ccccc@{}}
  \toprule
  Platform & Type & Cores & Process (nm)& Clock (MHz) \\ \midrule
  Xeon E5-2620 & CPU & 6 & 32 & 2000 \\
  Cortex A15 & CPU & 4 & 28 & 2300 \\
  Tegra Jetson K1 & GPU & 192 & 28 & 852 \\
  Titan X & GPU & 3072  & 28 & 1075 \\ 
  Kintex-7 & FPGA & N/A & 28 & 185 \\ 
  Automata Processor & AP & 64 & 50 & 133 \\ \bottomrule
\end{tabular}
\label{evaluated-platforms}
\end{table}

\subsection{Workload Parameters}

The run time of our NFA design scales with dimensionality so we use different feature vector lengths.
Practical vector dimensionality for feature descriptors like word embeddings~\cite{document-distances}, SIFT~\cite{sift}, and semantic embeddings~\cite{tagspace} range from 64 to 256; higher dimensional feature descriptors like those produced by AlexNet~\cite{alexnet} can be reduced using techniques like principle component analysis~\cite{pca}.
Past literature also shows the number of nearest neighbors $k$ ranges anywhere between 1 (exclusively nearest neighbor) and 20~\cite{sift, document-distances, motion-planning, how-many-results}.
The parameter sets we choose to evaluate are shown in Table~\ref{knn-parameters} for 4096 queries.

\begin{table}[t!]
  \centering
  \caption{kNN workload parameters.}
  \begin{tabular}{@{}ccc@{}}
    \toprule
    Workload & Dimensionality & Neighbors \\ \midrule
    kNN-WordEmbed  & 64             & 2        \\
    kNN-SIFT        & 128            & 4        \\
    kNN-TagSpace    & 256            & 16        \\ \bottomrule
  \end{tabular}
  \label{knn-parameters}
\end{table}

\subsection{Micron Automata Processor}

To estimate the performance of the AP, we simulate the design using the AP SDK 1.7.26 since full AP driver stacks at the time of writing were not available.
We implemented our designs in ANML, validated our design using the AP Workbench, and compiled each design to obtain resource utilization and frequency estimates.
We assume that the host processing program can operate concurrently (non-blocking API calls) with the AP much like how a CUDA program offloads to GPUs.
Since datasets in kNN applications are typically static, we do not include compile times since compilation can be done offline.
To estimate power, we connected a power meter to a single rank AP, and measured static and load power consumption; dynamic power is computed as the average load minus average static power and scaled to reflect a four rank device.
To estimate energy consumption for the AP we take the simulated run time and multiply by the dynamic power.
We then apply linear scaling factors to normalize the 50 nm AP lithography~\cite{ap-technode} to match the 28 nm lithographies of competing baselines.

\subsection{CPU, GPU, and FPGA}

For CPU baselines, we use the Hamming distance implementation from the Fast Library for Approximate Nearest Neighbors (FLANN)~\cite{flann}.
For GPU baselines, we modify an off-the-shelf CUDA implementation~\cite{gpu-knn} that uses a 32-bit XOR and population count (POPCOUNT) operation instead of a 32-bit Euclidean distance operation.
Run time is recorded as wall clock time and dynamic power is measured using a power meter by taking the difference between the load and idle power.
Energy consumption is estimated by multiplying dynamic power by run time.
To benchmark indexing times for spatial data structures, we use custom implementations and an off-the-shelf ITQ LSH implementation~\cite{lshbox}.
For simplicity, we use four hash tables for LSH and four parallel kd-trees; for kd-trees and k-means each tree traversal checks one bucket of vectors.

For FPGAs, we implement an AXI4-Stream compliant fixed-function kNN accelerator in Verilog for a Xilinx Kintex-7-325T.
The accelerator architecture consists of a scratchpad, XOR/POPCOUNT distance unit, and hardware priority queue. 
It also processes multiple queries in parallel; data vectors are streamed through the core, once per batch of queries.
We use Vivado 2014.4 to synthesize, place, and route the design, and report the estimated post-placement and route power using worst case activity factors.
To estimate run time, we simulate the design using the Vivado simulator, and estimate power using the Vivado power analyzer.

\section{Results} \label{sec:evaluation}

\subsection{Resource Utilization}

We first evaluate resource utilization since the number of vectors that can be processed per board configuration will affect AP performance.
We report the total rectangular block area based on compilation reports generated by the \code{apadmin} tool.
For each of our parameter sets kNN-WordEmbed, kNN-SIFT, and kNN-TagSpace, we observe resource utilizations of 41.7\%, 90.9\%, and 78.6\% of the total resource capacity respectively~\footnote{For kNN-WordEmbed, we are limited by the theoretical peak PCIe bandwidth and thus cannot exploit all the resources.}.
This equates up to 128 Kb of encoded data per board configuration on current generation APs (1024$\times$128 dimensions or 512$\times$256 dimensions).
Finally, we note that resource utilization does not depend on the number of neighbors $k$ since sorting does not require additional automata states.

\subsection{Run Time Performance}

We now compare AP run time performance against competing solutions.
Run time performance comparisons are shown in Table~\ref{small-dataset} and Table~\ref{large-dataset} for a small dataset (512-1024 points) and large dataset ($2^{20} \approx$ 1 million points) respectively.
The small dataset illustrates the performance of the AP for kNN problems that fit on one AP board configuration. 
For the small dataset, the AP achieves an order of magnitude performance improvement over competing CPU solutions.
This performance improvement can be attributed to a combination of the parallel processing capacity of the AP, and temporally encoded sort which dramatically lowers the run time of the sorting step from $O(n\log n)$ to $O(d)$.
We observe poor GPU performance likely due to poor blocking of the binarized data; since vectors are now 32 times smaller (1 bit per dimension), the finer grained memory accesses are less optimal for the off-the-shelf baseline.
In contrast, the large dataset reflects the performance for kNN problems that exceed the capacity of a single AP board.
For the large dataset, AP Gen 1 performance degrades due to dynamic reconfiguration overheads which account for upwards of 98\% of the execution time.
To quantify the impact of this bottleneck, we provide a second estimate with AP Gen 2 reconfiguration latencies ($\sim$100$\times$ better).
Our results show that the reconfiguration time improvement yields 19.4$\times$ performance improvement between Gen 1 and Gen 2, and shifts the bottleneck back to the computation.
Finally, we also show the projected performance of the AP with the automata optimizations and architectural extensions presented in Section~\ref{sec:optimizations} and~\ref{sec:analysis} (AP Opt+Ext).

{
\setlength{\belowcaptionskip}{-10pt}
\begin{table}[!t]
\centering
\caption{kNN Performance and Energy Efficiency for Small Datasets.}
\label{small-dataset}
\begin{tabular}{@{}cccccc@{}}
\toprule
\multicolumn{6}{c}{\textbf{Run Time Performance (ms) - Lower is Better}} \\ \midrule
kNN-Workload & \begin{tabular}[c]{@{}c@{}}Xeon\\ E5-2620\end{tabular} & \begin{tabular}[c]{@{}c@{}}Cortex\\ A15\end{tabular} & \begin{tabular}[c]{@{}c@{}}Jetson\\ TK1\end{tabular} & \begin{tabular}[c]{@{}c@{}}Kintex\\ 7\end{tabular} & \begin{tabular}[c]{@{}c@{}}AP\\ Gen 1\end{tabular} \\
WordEmbed ($n$=1024) & 23.33 & 103.63 & 125.80 & 1.89 & 1.97 \\
SIFT ($n$=1024) & 37.50 & 191.44 & 155.94 & 3.78 & 3.94 \\
TagSpace ($n$=512) & 33.97 & 185.34 & 160.15 & 4.33 & 7.88 \\ \midrule
\multicolumn{6}{c}{\textbf{Energy Efficiency (query / Joule) - Higher is Better}} \\ \midrule
kNN-Workload & \begin{tabular}[c]{@{}c@{}}Xeon\\ E5-2620\end{tabular} & \begin{tabular}[c]{@{}c@{}}Cortex\\ A15\end{tabular} & \begin{tabular}[c]{@{}c@{}}Jetson\\ TK1\end{tabular} & \begin{tabular}[c]{@{}c@{}}Kintex\\ 7\end{tabular} & \begin{tabular}[c]{@{}c@{}}AP\\ Gen 1\end{tabular} \\
WordEmbed ($n$=1024) & 3344 & 4941 & 27133 & 579214 & 110445 \\
SIFT ($n$=1024) & 2081 & 2674 & 21889 & 289607 & 44603 \\
TagSpace ($n$=512) & 2297 & 2762 & 21314 & 253406 & 22301 \\ \bottomrule
\end{tabular}
\end{table}

}

{
\captionsetup{skip=-10pt}

\begin{table*}[!t]
\centering
\caption{kNN Performance and Energy Efficiency for Large Datasets.}
\label{large-dataset}
\begin{tabular}{@{}ccccccccc@{}} \\ \toprule
\multicolumn{9}{c}{\textbf{Run Time Performance (s) - Lower is Better}} \\ \midrule
Workload & Xeon E5-2620 & Cortex A15 & Jetson TK1 & Titan X & Kintex 7 & AP Gen 1 & AP Gen 2 & AP (Opt+Ext) \\
WordEmbed & 19.89 & 109.06 & 16.09 & 0.99 & 1.85 & 48.10 & 2.48 & 0.039 \\
SIFT & 33.18 & 199.5 & 16.73 & 1.02 & 3.69 & 50.11 & 4.50 & 0.062 \\
TagSpace & 60.12 & 382.82 & 16.41 & 1.03 & 7.38 & 108.31 & 17.07 & 0.23 \\ \midrule
\multicolumn{9}{c}{\textbf{Energy Efficiency (query / Joule) - Higher is Better}} \\ \midrule
Workload & Xeon E5-2620 & Cortex A15 & Jetson TK1 & Titan X & Kintex 7 & AP Gen 1 & AP Gen 2 & AP (Opt+Ext) \\
WordEmbed & 3.92 & 4.69 & 212.14 & 83.84 & 593.89 & 4.53 & 87.81 & 1737.92 \\
SIFT & 2.35 & 2.57 & 204.02 & 81.94 & 296.95 & 4.34 & 48.40 & 1091.86 \\
TagSpace & 1.30 & 1.34 & 208.00 & 81.05 & 148.47 & 1.62 & 10.20 & 236.30 \\ \bottomrule
\end{tabular}
\end{table*}
}

We now evaluate and compare the efficacy of spatial indexing techniques relative to equivalent CPU baselines.
We use an analytical model to estimate run time by benchmarking the index traversals on the CPU, and adding it to estimated AP reconfiguration and simulated run time.
In our model, we batch searches to the same bucket where possible and report the average run time over several iterations; each bucket size is set to the capacity of one AP board configuration (512-1024 points).
Decreasing bucket sizes further would not yield run time improvements, underutilize the AP, and unnecessarily lower search accuracy since less data is scanned.
Table~\ref{spatial-indexing} shows the relative performance of the AP for indexing techniques compared to single threaded CPU baselines for a large kNN-TagSpace workload.
For indexing, we observe poorer performance for AP Gen 1 because reconfiguration time dominates run time.
For for Gen 2, we observe good run time improvements since run time is now dominated by the linear bucket scan.

{

\begin{table}[!ht]
\centering
\caption{Relative speeds up for spatial indexing techniques on kNN-TagSpace.}
\begin{tabular}{@{}cccc@{}} 
  \toprule
  Indexing & ARM + AP Gen 1 & ARM + AP Gen 2 \\ \midrule
  Linear (No Index) & 16$\times$ & 91$\times$ \\ 
  KD-Tree & 0.89$\times$ & 106$\times$ \\ 
  K-Means & 0.88$\times$ & 120$\times$ \\ 
  MPLSH   & 0.62$\times$ & 3.5$\times$ \\ \bottomrule
\end{tabular}
\label{spatial-indexing}
\end{table}
}

\subsection{Energy Efficiency}

The estimated energy efficiency of the AP for the small and large datasets are shown in Table~\ref{small-dataset} and Table~\ref{large-dataset} respectively.
Our results show that the Gen 1 AP can provide up to 43$\times$ energy efficiency improvement over general purpose cores and is largely in line with the run time improvements we observe.
When compared to production FPGA and GPU platforms, our results show that the Gen 1 AP is again severely outperformed due to high reconfiguration overheads. 
However, with Gen 2 reconfiguration times the AP achieves competitive energy efficiency and is within one order of magnitude of competing production platforms.
With architectural extensions and design optimizations on next generation devices, we expect the AP could potentially surpass FPGA fabrics in energy efficiency. 

\section{Automata Optimization Techniques}
\label{sec:optimizations}

We now introduce three mutually orthogonal automata design optimizations: \textit{vector packing}, \textit{symbol stream multiplexing}, and \textit{statistical activation reduction}.

\subsection{Vector Packing}

To more efficiently utilized the STE resources, we propose \textit{vector packing}.
The key insight is that Hamming macros share common portions which can be combined by overlaying them over the same base NFA to reduce resource utilization.
Fig.~\ref{vector-packing} shows how to pack Hamming macros encoding the vectors $\{1, 1, 0, 1\}$ (Fig.~\ref{vector-packing}a) and $\{1, 0, 0, 0\}$ (Fig.~\ref{vector-packing}b) into a single NFA (Fig.~\ref{vector-packing}c).
We refer to the boxed portion in Fig.~\ref{vector-packing}c as a \textit{vector ladder} that is the base NFA used to overlay vectors on.
For each vector packed, we instantiate extra collector states, and sorting macros and connect them to the appropriate states in the vector ladder.
In theory, with unlimited routing, all dataset vectors can be packed into a single NFA but in practice we are limited by the maximum state in-degree, and NFA size.
So instead, vectors would be packed until we reach one of these limitations; we would then start a new NFA and repeat until all vectors are processed.

We evaluate vector packing by running a microbenchmark which places and routes eight vectors across 32, 64, and 128 dimensions, and compare their resource utilizations.
Surprisingly, we find that vector packing is ineffective in practice at reducing the resource footprint when compared to the theoretical gains projected by a simple analytical model.
We believe this is due to the increased routing pressure which imposes reduced compilation feasibility.
For high dimensional designs, we find that the compiler has difficulty routing the design leading to placed but only partially routed compilations.
The routing pressure is most likely due to the vector ladder which introduces many states with high fan outs.
However, we expect that in future generation APs and with tool chain maturity, that vector packing will eventually provide viable improvements.
To estimate the potential improvements, we build a simple analytical model where each NFA state incurs one STE resource cost.
The maximum potential savings is then computed as the resource cost of the original NFA over the resource cost of the vector packed NFA.

{
\setlength{\belowcaptionskip}{-10pt}
\begin{figure*}[!ht]
\centering
\includegraphics[width=\linewidth]{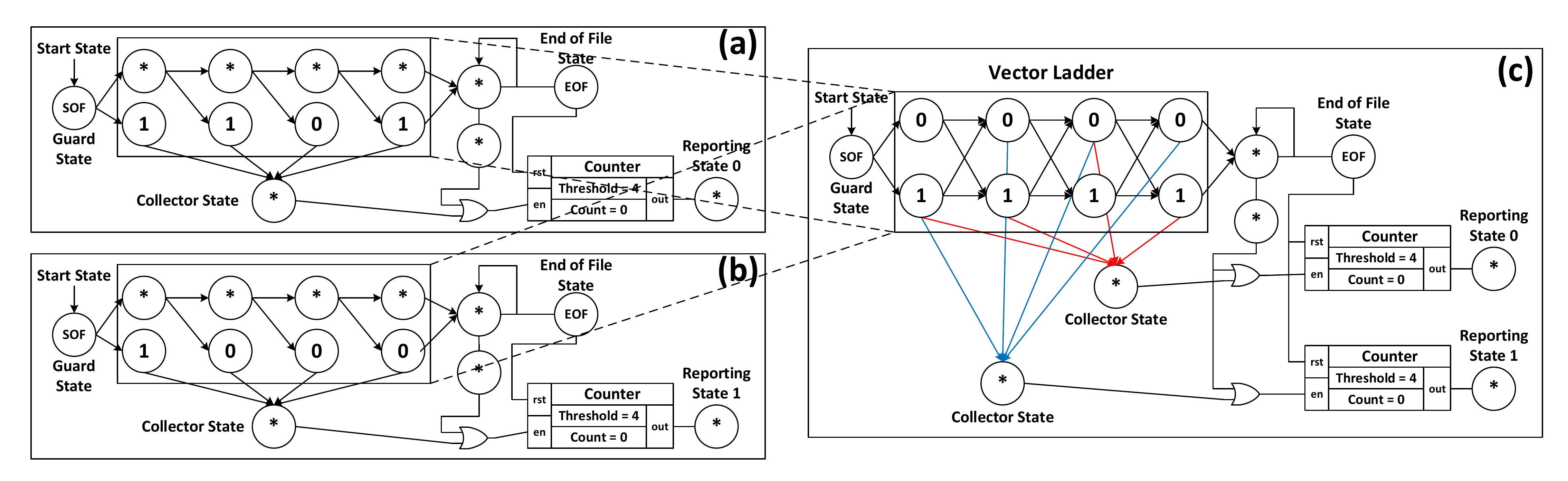}
\caption{Vector packing for two distinct Hamming macros. (a) Hamming macro encoding \{1, 1, 0, 1\}. (b) Hamming macro encoding \{1, 0, 0, 0\}. (c) Combined NFA encoding both vectors.} 
\label{vector-packing}
\end{figure*}
}

\subsection{Symbol Stream Multiplexing}

Another inefficiency in our automata design is that each state in the vector ladder only effectively uses one bit of information in the symbol stream (the symbol is either 0 or 1).
However, symbol streams are 8 bits wide so processing a single query vector at a time is wasteful.
To improve the throughput of the design, the unused bits in the symbol stream can encode additional parallel query vectors which can provide up to an 7$\times$ throughput improvement by exploiting query level parallelism. 
To support parallel queries, up to seven parallel NFAs for each dataset vector can be instantiated where each NFA PCRE is programmed to process a different bit slice of the symbol stream.
We cannot achieve an 8$\times$ improvement because of special symbols like the SOF and EOF.

To discriminate among the bits in the symbol, the STEs can be programmed to effectively perform a ternary match similar to how ternary matches are encoded on TCAMs. 
This can be achieved by exhaustively encoding all extended ASCII characters which satisfy the ternary match.
For instance, an NFA which matches the first bit in the symbol as 1 would require a PCRE corresponding with the ternary match \code{0b*******1} (where \code{*} denotes a ternary match).
Similarly, the PCRE for a STE that matches 0 as the first bit would be \code{0b*******0}
A similar set of ternary matches can be constructed to discriminate among other bit slices in the stream.
Finally, symbol stream multiplexing is orthogonal to vector packing since it only requires duplicating existing NFA structures and replacing STE symbols.

\begin{figure}[!ht]
\centering
\includegraphics[width=\linewidth]{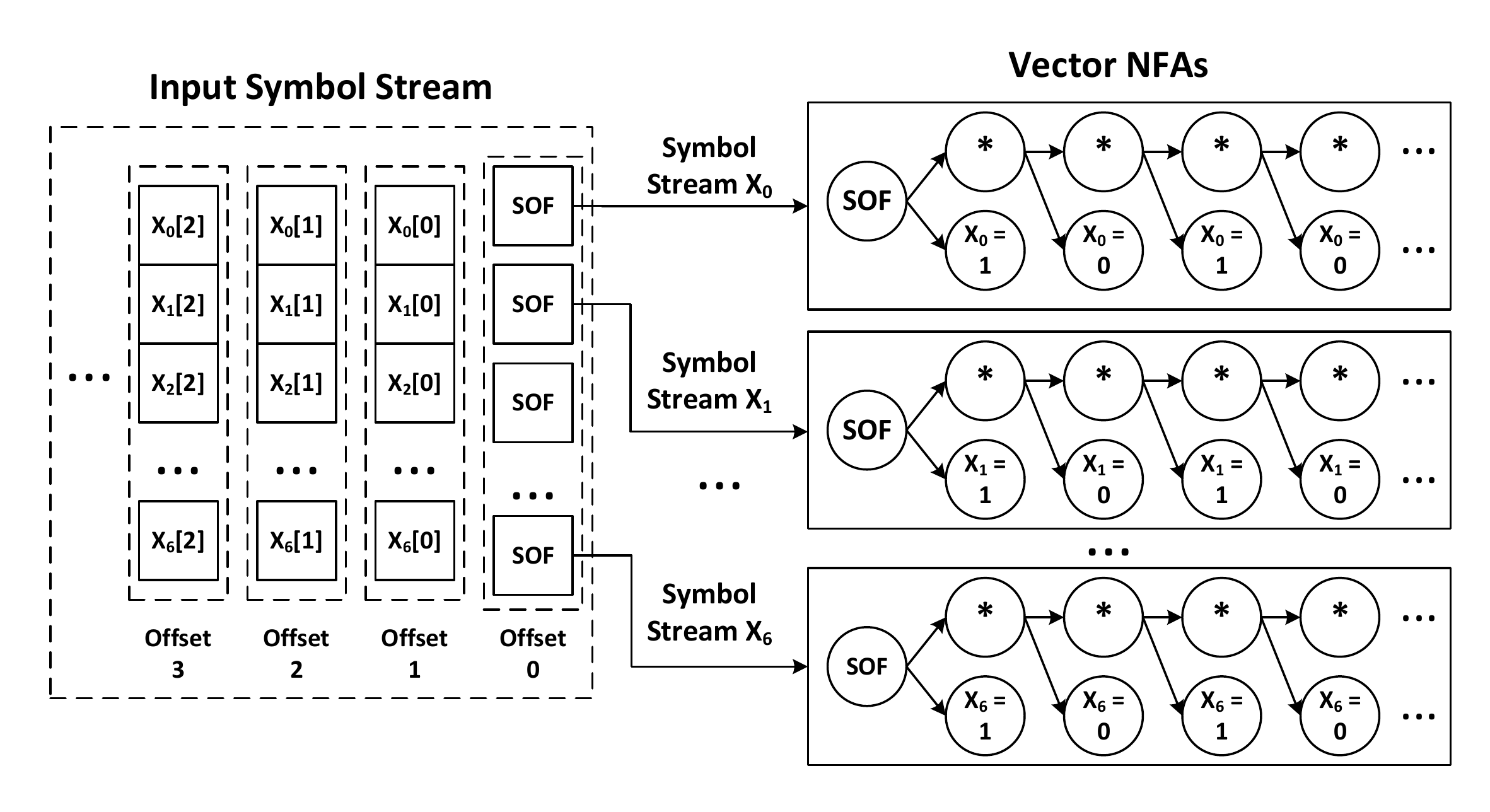}
\caption{Symbol stream multiplexing: NFA STEs are replicated and encoded to discriminate among different bit slices of the symbol stream.}
\label{stream-slicing}
\end{figure}

For the current generation AP, there is neither sufficient resources, nor sufficient external bandwidth to implement stream multiplexing.
Replicating the base design 7$\times$ is infeasible since our design already uses 41-91\% of the board capacity.
Similarly, replicating the number of reporting states 7$\times$ requires in excess of 200 Gbps of external PCIe bandwidth to convey all the activations.
However, with future technology scaling, higher board capacities, better PCIe interconnects, and statistical reduction (discussed next), stream multiplexing should become viable.

\subsection{Statistical Activation Reduction}

As the sorting step completes, the AP reports state activations to the host processing unit.
We assume the AP communicates state activations by using a bit mask and reporting an offset; the offset indicates which cycle in the symbol stream the activations correspond to.
Since activations are often sparse, the bit mask can be compressed to use a sparse vector encoding (with 32-bit offsets). 
For a kNN design that encodes $n$ vectors on the AP, we expect $32 \times n$ bits to convey the activations plus $32 \times d$ bits to convey the activation time step.
A single query (which has a latency of $2d$ cycles), requires the AP to convey $32 \times (n + d)$ bits every $2d \times 7.5 ns$ (133 MHz design).
This amounts to 36.2, 18.1, and 9.0 Gbps of sustained external bandwidth for kNN-WordEmbed, kNN-SIFT, and kNN-TagSpace respectively which are significant fractions of the PCIe Gen 3$\times$ 8 bandwidth (63 Gbps).

Ideally, we want to suppress state activations after the top $k$ results from the temporal sort have been communicated.
However, dynamically inserting an EOF symbol to reset the automata is not possible (Section~\ref{sec:ap-description}), and building an NFA to generate a global reset signal would couple all NFAs and exceed the maximum NFA size.
So instead, we propose a combination of local reductions and statistical approximations to suppress additional state activations while still providing mostly correct results.

This can be done by first partitioning vector NFAs into groups of $p$ NFAs. 
For a dataset $D$ this yields $R = |D|/p$ NFA partitions.
For each partition, we use a counter to track how many reporting state activations have occurred locally.
If the global number of desired neighbors is $k$, for each group of Hamming macros we only report the top $k'$ results where $k' < k$ and $k' \times R > k$.
This is can be done by resetting all of the inverted Hamming distance counters after $k'$ reporting states have activated within a set of macros shown in Fig.~\ref{reduction-macro}.
Each of the $k'$ local result sets can then be sorted globally on the host processing unit.
The key insight is that instead of reporting all $p$ state activations per set, we only need to report $k'$ states which provides a bandwidth reduction by a factor of $p/k'$ and still maintain mostly correct results.
For appropriately set values of $k'$, $k$, and $p$, the probability that the global result set does not contain the global top $k$ neighbors can be made arbitrarily small.
Even for failing cases, the result sets returned by this approximation will be mostly correct.

We use a statistical model to determine how much data reduction is achieved without compromising accuracy.
For each configuration, we randomly generate dataset and query vectors, partition the dataset vectors, execute local kNN, and perform global top-$k$ sort to determine if exact kNN results are computed.
We repeat the process 100 times to determine statistically how often an incorrect result is returned.
Table~\ref{statistical-model} shows results for different values of $k'$ and $k$ used in our evaluation.
Our results show that for reasonable parameters we can achieve good bandwidth reductions and still achieve acceptable result accuracy.
Finally, we note this optimization is synergistic with vector packing since vector packing provides a natural grouping of vectors.

{

\begin{figure}[!t]
\centering
\includegraphics[width=\linewidth]{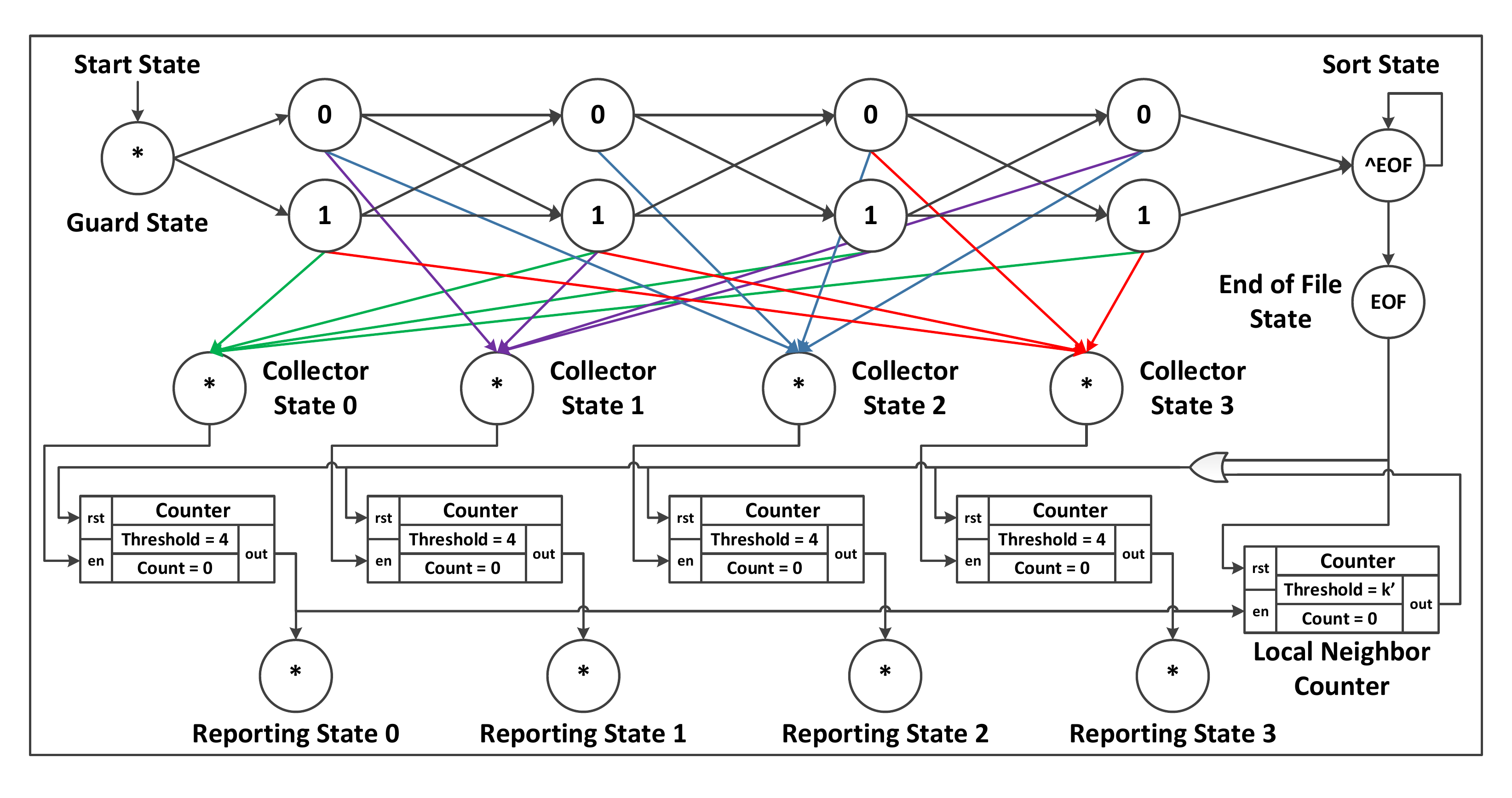}
\caption{Automata for four vectors with reduction states and counter. The Local Neighbor Counter resets all other counters suppressing unnecessary activations.\protect\footnotemark}
\label{reduction-macro}
\end{figure}

}
\footnotetext{Edges from sorting macro to counter enables not shown for clarity.}
{
\captionsetup{skip=-10pt}

\begin{table}[!t]
\centering
\caption{Percentage of incorrect results out 100 randomized runs for $p=16$, $n=1024$ (lower is better).} 
\begin{tabular}{@{}ccccccccc@{}} \\ \toprule
Workload & $k$  & $k'=1$ & $k'=2$ & $k'=3$ & $k'>= 4$ \\ \midrule
WordEmbed & 2  & 100\% & 1\% & 0\% &  0\% \\
SIFT      & 4  & 100\% & 1\% & 0\% &  0\% \\
TagSpace  & 16 & 100\% & 72\% & 5\% &  0\% \\ \bottomrule
\end{tabular}
\label{statistical-model}
\end{table}
}

\section{Architecture Extension Recommendations}
\label{sec:analysis}

We now present microarchitectural extensions and briefly evaluate their resource utilization and performance impact on future AP hardware.

\subsection{Counter Increment Extension}

One limitation of the AP is the lack of arithmetic units which limits arithmetic calculations to those that increment by at most one (e.g., Hamming or Jaccard distance).
In the context of kNN, allowing the counters to take multiple increment signals would enable processing of multiple vector dimensions in parallel per time step by encoding up to 7 dimensions per symbol.
Applying this optimization would preclude stream multiplexing for kNN since it uses the higher symbol bits to encode dimensions of the same query instead of the same dimension across queries.
This optimization would reduce the query latency by 7$\times$ but leave the sorting latency unchanged.
Quantitatively, since the Hamming macro and sorting macro each take $d$ cycles to execute, the counter increment extension reduces the query latency to $d + d/7$ cycles which is a 43\% improvement or 1.75$\times$ better.

More generally, this also reduces the need for collector states when performing reductions or counting multiple activations in parallel.
For example, this modification can also enhance the throughput and latency of the apriori frequent item set mining automata presented in~\cite{ap-apriori} to increase the number of dataset vectors processed in parallel for a given itemset.
Modifying the 8-bit counters to increment by up to 8 can efficiently be done with a carry save adder but requires 2.8$\times$ more two-input gates.
However, since the AP fabric contains roughly two orders of magnitude fewer counters than STEs this modification should incur minimal overall area overhead.

\subsection{Counter Dynamic Threshold Extension}

\begin{figure}[!th]
\centering
\includegraphics[width=\linewidth]{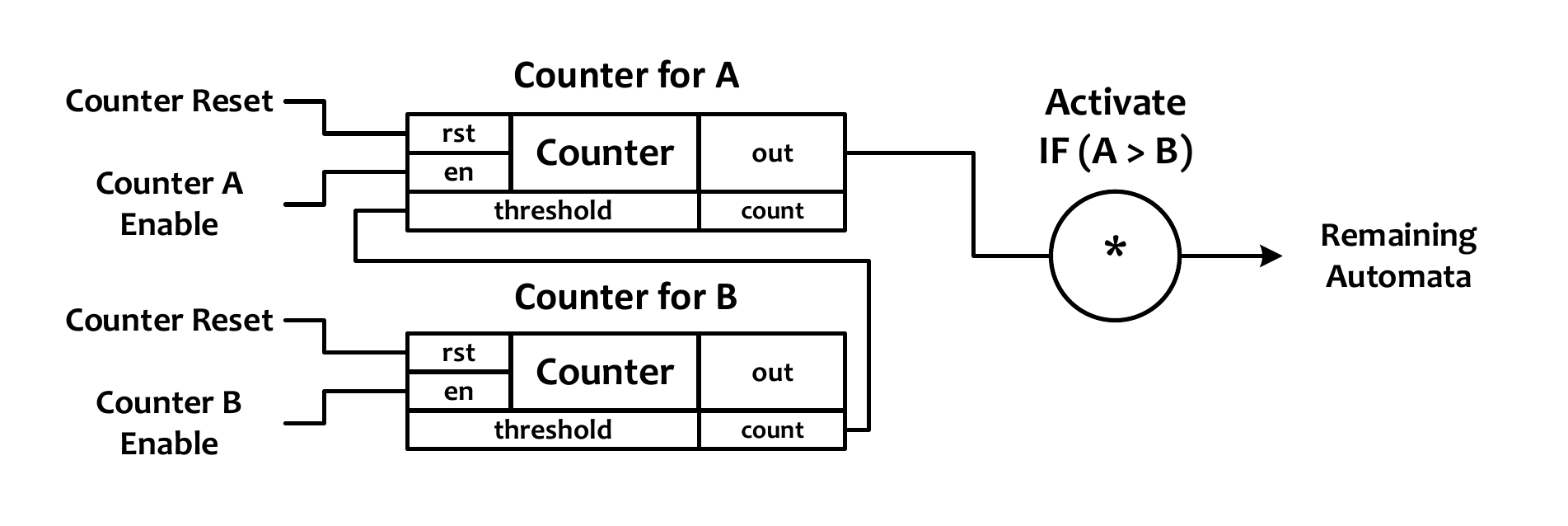}
\caption{Comparison macro using dynamically thresholded counters.}
\label{comparison-automata}
\end{figure}

A second counter extension is to allow dynamic counter thresholds; currently counter thresholds are fixed at design time which eliminates any way of implementing data dependent dynamic comparisons and control structures.
A simple extension is to expose a reconfigurable counter threshold port which can be driven by the internal count of another counter enabling computations that require dynamically computed information.
Most notably this modification can be used to build automata constructs that require arithmetic comparisons between two dynamic values as shown in Fig.~\ref{comparison-automata}.
This automata can be used to build a more powerful ``if (A $>$ B) ... else ...'' construct which is more useful than the currently available ``if (A $>$ threshold) ... else ...''.
This extension requires no extra hardware resources and only requires a few wires in the routing fabric to accommodate newly exposed internal counts.

\subsection{STE Decomposition Extension}

Another key challenge of the AP fabric is fully utilizing the symbol address space of the STEs.
For instance, in the context of kNN, only one bit of information is used by each STE per symbol to determine the state output.
This means our kNN automata design underutilizes an 8-input STE or lookup table (LUT) effectively as a 1-input LUT since the remaining 7 bits of the input are not used.
In terms of resource usage this is wasteful; a simple extension is to allow decomposition of STEs  into multiple smaller STEs, or reducing the STE size so that they can be composed when larger STEs are needed.

\begin{figure}[!t]
\centering
\includegraphics[width=0.9\linewidth]{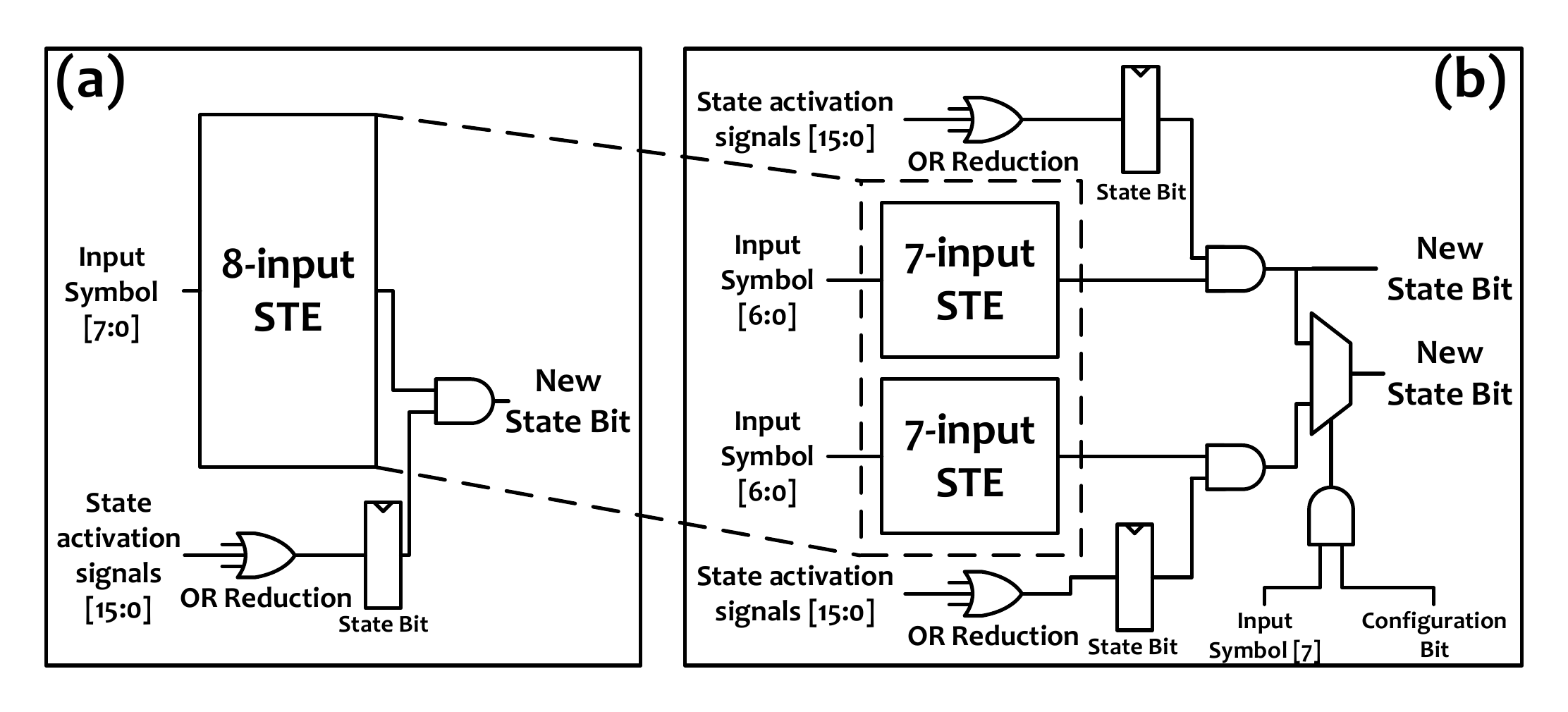}
\caption{STE decomposition from a 8-LUT (a) to two 7-LUTs (b).}
\label{ste-decomposition}
\end{figure}

The key insight is that an 8-input LUT can easily be composed from two smaller 7-input LUTs (Fig.~\ref{ste-decomposition}), which can each be further decomposed if necessary.
We refer to the degree to which an 8-input STE is decomposed as the \textit{decomposition factor} $x$; for instance, if we use an 8-input LUT as 4 $\times$ 6-input LUTs, the decomposition factor is $x = 4\times$.
By performing this decomposition, more NFA states can be packed into roughly the same number of 8-input STE resources; this is because states in the NFA that can be implemented using smaller lookup tables or STEs can be packed together.
To evaluate this extension, we build an analytical model to estimate savings in the context of kNN. 
In our model we assume each state in the original automata has a resource cost of one 8-input STE.
We then estimate how many 8-input STEs would be required if different degrees of STE decomposition are implemented.
For instance, if an 8-input STE can be decomposed into 2 $\times$ 7-input STEs, three states which each only need to see 6 bits of the symbol can be packed into two 8-input STEs incurring a cost of two in our analytical model.
Similarly, these three states can be packed into a single 8-input STE if it can be broken down into 4 $\times$ 6 input STEs incurring a cost of one.
Since the Hamming macro portion of the kNN automata dominates the STE cost, our model shows STE decomposition can potentially provide close to linear reductions in STE resources (Table~\ref{ste-decomposition-resources}).

More generally, STE decomposition can improve resource usage for generalized regular expressions.
It is not unreasonable to assume that the extended ASCII characters (ASCII values 128-255) frequently remain unused in many PCREs.
Since the original ASCII character set consists of only 128 symbols, two 7-bit STEs can be encoded in a decomposed 8-bit STE yielding 2$\times$ resource utilization improvement.
Practically there is a limit to which STEs should be decomposed (e.g., FPGAs use 6-LUTs) since decomposing STEs indefinitely incurs increased circuit control overheads. 

{
\captionsetup{skip=-10pt}

\begin{table}[!b]
\centering
\caption{STE decomposition resource savings.}
\begin{tabular}{@{}ccccccc@{}} \\ \toprule
\begin{tabular}[c]{@{}c@{}}Decomposition\\Factor\end{tabular} & $x=1$ & $x=2$ & $x=4$ & $x=8$ & $x=16$ & $x=32$  \\ \midrule
WordEmbed & 1$\times$ & 1.98$\times$ & 3.86$\times$ & 7.38$\times$ & 13.56$\times$ & 23.34$\times$ \\
SIFT & 1$\times$ & 1.99$\times$ & 3.93$\times$ & 7.67$\times$ & 14.68$\times$ & 27.00$\times$ \\
TagSpace & 1$\times$ & 1.99$\times$ & 3.96$\times$ & 7.83$\times$ & 15.31$\times$ & 29.26$\times$ \\
Theoretical & 1$\times$ & 2$\times$ & 4$\times$ & 8$\times$ & 16$\times$ & 32$\times$ \\ \bottomrule
\end{tabular}
\label{ste-decomposition-resources}
\end{table}
}

\subsection{Summary}

Resource savings on the AP translates directly to performance improvements since more parallel NFAs can be executed in parallel.
By compounding the mutually orthogonal architectural extensions with automata design optimizations, we estimate that APs can achieve \textit{up to an additional} 73$\times$ potential performance improvement over Gen 2 results.
The additional compute density from technology scaling incurs power overheads so we expect energy efficiency to only improve by up to 23$\times$.
Table~\ref{summary-table} summarizes the projected improvements assuming natural technology scaling from 50nm to 28nm, 4$\times$ decomposition factor, vector packing into groups of 4, and 8-input counter increment extensions.
Performance improvement yields from vector packing and STE decomposition are calculated using analytical models based on the number of states saved in the original NFA (1 NFA state $\approx$ 1 STE resource).
For STE decomposition, we assume a decomposition factor of 4 which effectively breaks 8-input STEs into 6-input STEs; this is roughly equivalent to the 6-LUTs used in FPGA fabrics.
With these additional improvements, \textit{the AP has the potential to be comparable to or even better than FPGAs and GPUs} for the kNN workloads presented in this paper.

{

\begin{table}[!ht]
\centering
\caption{Total compounded \textit{additional} potential performance gains from optimizations and architectural extensions.}
\label{summary-table}
\begin{tabular}{@{}cccc@{}} \\ \toprule
Workload& kNN-WordEmbed& kNN-SIFT& kNN-TagSpace \\ \midrule
Technology Scaling&     3.19$\times$& 3.19$\times$& 3.19$\times$ \\
Vector Packing&         2.93$\times$& 3.28$\times$& 3.31$\times$ \\
STE Decomposition&      3.86$\times$& 3.93$\times$& 3.96$\times$ \\
Counter Increment Ext.& 1.75$\times$& 1.75$\times$& 1.75$\times$ \\
\textbf{Total Improvement}& \textbf{63.14$\times$}&\textbf{71.96$\times$}& \textbf{73.17$\times$} \\ \bottomrule
\end{tabular}
\end{table}
}

\section{Related Work} \label{sec:related-work}

We are not the first to propose accelerating kNN using alternative computing substrates or near-data processing techniques. 
As far back as the late 1980s, Kanerva et al.~\cite{sparse-distributed-memory} proposed Sparse Distributed Memory (SDM) for nearest neighbor search.
This was followed by Roberts et al.~\cite{pcam} who proposed proximity content-addressable memory (PCAM) which employed the ideas behind SDM.
Lipman et al.~\cite{smart-access-memory} then proposed Smart Access Memory which integrated distance calculation units with a systolic array sort for kNN.
Finally, Tandon et al.~\cite{tandon13} proposed an all-pairs similarity search accelerator extension for the L2-cache to accelerate NLP applications.

More recently, work has shown it is possible to implement approximate variants of kNN using TCAMs.
Most notably, Shinde et al.~\cite{shinde10} propose ternary locality sensitive hashing (TLSH) and Bremler-Barr et al.~\cite{bremler-barr15} propose using binary-reflected Gray code to implement approximate nearest neighbors.
Unfortunately the accessibility of TCAMs has traditionally been limited to niche hardware applications (e.g., routers) making them difficult to program, access, and evaluate.

To our knowledge, we are the first to take advantage of temporal encodings in NFAs to accelerate automata calculations for kNN on the AP.
Past work in the AP application space has focused mainly on pattern mining applications like biological motif search~\cite{ap-motif-search}, frequent itemset mining~\cite{ap-apriori}, and graph pattern mining~\cite{ap-graph}.
There have also been efforts at porting decision tree applications like random forests~\cite{ap-random-forest} to the AP.
The most relevant work to ours in this space is by Ly et al.~\cite{ly16} who propose automata-based feature extraction and image retrieval encodings.
However, their proposed automata only supports exact categorical matches for image retrieval while our work implements kNN which is significantly more general.
Thus, our work augments the AP application space by introducing a novel similarity search automata design which we have shown is both fast and energy efficient.

Finally, application-driven PIM accelerators are not a new idea but have enjoyed renewed interest with the advent of new technologies like die-stacked memory; recent proposals have also ranged from fully programmable solutions, to domain specific accelerator fabrics, to fixed-function accelerators.
Examples of fully programmable solutions can be found in work by Zhang et al.~\cite{zhang14} and Hsieh et al.~\cite{hsieh16} which both propose integrating GPUs with 3D memory to alleviate bandwidth bottlenecks.
Farmahini-Farahani et al.~\cite{farmahini-farahani15} also propose a PIM architecture which integrates a coarse-grained reconfigurable array on top of commodity DRAMs; tasks were then offloaded to the array via a general purpose processor.
Instances of more domain specific PIM proposals include work by Ahn et al.~\cite{ndp-graph} who propose a PIM architecture for parallel graph processing, and Xie et al.~\cite{pim-graphics} who propose PIM to accelerate 3D graphics rendering on top of Hybrid Memory Cube.
Our work is another instantiation of the general trends towards exploiting the benefits of application-driven PIM offload using the AP.

\section{Conclusion} \label{sec:conclusion}

We presented a novel automata design and several optimization techniques for similarity search which exploits a temporally encoded sort for automata processors.
Our results show that our automata design can achieve significant performance gains at comparable energy efficiency as competing CPU solutions.
We also proposed architectural modifications and NFA design optimizations that expose the latent potential of the AP for similarity search.
We expect that with better technology scaling, and improved reconfiguration latency that the AP will be able to support significantly larger workloads.
We also expect that as the platform matures the architecture, resource capacity, and energy efficiency will improve dramatically making the AP a promising platform for many other automata-based applications.

\section{Acknowledgements}

We would like to thank all of our anonymous reviewers and colleagues for their valuable and insightful feedback.
We especially thank Matt Grimm, Matt Tanner, and Indranil Roy at Micron Technologies Incorporated for taking the time to help guide, and validate our designs, and provide power measurements on actual AP hardware.
This work was also supported in part by NSF under grant CCF-1518703, gifts by Oracle, and by C-FAR, one of the six SRC STARnet Centers, sponsored by MARCO and DARPA.


\balance
\bibliographystyle{ieeetr}
\bibliography{references}

\begin{thebibliography}{10}

\bibitem{document_distances}
M.~Kusner, Y.~Sun, N.~Kolkin, and K.~Q. Weinberger, ``{"From Word Embeddings To
  Document Distances"},'' in {\em Proceedings of the 32nd International
  Conference on Machine Learning (ICML-15)} (D.~Blei and F.~Bach, eds.),
  pp.~957--966, JMLR Workshop and Conference Proceedings, 2015.

\bibitem{content_based_search}
Y.~Liu, D.~Zhang, G.~Lu, and W.-Y. Ma, ``{A survey of content-based image
  retrieval with high-level semantics},'' {\em Pattern Recognition}, vol.~40,
  no.~1, pp.~262 -- 282, 2007.

\bibitem{dedup}
L.~Aronovich, R.~Asher, E.~Bachmat, H.~Bitner, M.~Hirsch, and S.~T. Klein,
  ``The design of a similarity based deduplication system,'' in {\em
  Proceedings of SYSTOR 2009: The Israeli Experimental Systems Conference},
  SYSTOR '09, (New York, NY, USA), pp.~6:1--6:14, ACM, 2009.

\bibitem{facebook-photos}
D.~Beaver, S.~Kumar, H.~C. Li, J.~Sobel, and P.~Vajgel, ``Finding a needle in
  haystack: Facebook's photo storage,'' in {\em Proceedings of the 9th USENIX
  Conference on Operating Systems Design and Implementation}, OSDI'10,
  (Berkeley, CA, USA), pp.~47--60, USENIX Association, 2010.

\bibitem{youtube-statistics}
YouTube, ``{Statistics - YouTube},'' 2014.

\bibitem{rebooting-the-it-revolution}
``Rebooting the it revolution: A call to action,'' 2015.

\bibitem{google-images}
T.~Liu, C.~Rosenberg, and H.~Rowley, ``{Clustering billions of images with
  large scale nearest neighbor search},'' 2007.

\bibitem{facebook-search}
Q.~Wang, D.~Zhang, and L.~Si, ``Weighted hashing for fast large scale
  similarity search,'' in {\em Proceedings of the 22nd ACM International
  Conference on Information \& Knowledge Management}, CIKM '13, (New York, NY,
  USA), pp.~1185--1188, ACM, 2013.

\bibitem{McKee95}
W.~A. Wulf and S.~A. McKee, ``Hitting the memory wall: Implications of the
  obvious,'' {\em SIGARCH Comput. Archit. News}, vol.~23, pp.~20--24, Mar.
  1995.

\bibitem{iram}
D.~Patterson, T.~Anderson, N.~Cardwell, R.~Fromm, K.~Keeton, C.~Kozyrakis,
  R.~Thomas, and K.~Yelick, ``A case for intelligent ram,'' {\em IEEE Micro},
  vol.~17, pp.~34--44, Mar. 1997.

\bibitem{flexram}
Y.~Kang, W.~Huang, S.-M. Yoo, D.~Keen, Z.~Ge, V.~Lam, P.~Pattnaik, and
  J.~Torrellas, ``Flexram: toward an advanced intelligent memory system,'' in
  {\em Computer Design, 1999. (ICCD '99) International Conference on},
  pp.~192--201, 1999.

\bibitem{activepages}
M.~Oskin, F.~T. Chong, and T.~Sherwood, ``Active pages: A computation model for
  intelligent memory,'' {\em SIGARCH Comput. Archit. News}, vol.~26,
  pp.~192--203, Apr. 1998.

\bibitem{ap-motif-search}
I.~Roy and S.~Aluru, ``Discovering motifs in biological sequences using the
  micron automata processor,'' {\em IEEE/ACM Trans. Comput. Biol.
  Bioinformatics}, vol.~13, pp.~99--111, Jan. 2016.

\bibitem{ap-apriori}
K.~Wang, Y.~Qi, J.~J. Fox, M.~R. Stan, and K.~Skadron, ``Association rule
  mining with the micron automata processor,'' in {\em Proceedings of the 2015
  IEEE International Parallel and Distributed Processing Symposium}, IPDPS '15,
  (Washington, DC, USA), pp.~689--699, IEEE Computer Society, 2015.

\bibitem{ap-graph}
I.~Roy, {\em Algorithmic Techniques for the Micron Automata Processor}.
\newblock PhD thesis, Georgia Institute of Technology, 2015.

\bibitem{document-distances}
M.~Kusner, Y.~Sun, N.~Kolkin, and K.~Q. Weinberger, ``{"From Word Embeddings To
  Document Distances"},'' in {\em Proceedings of the 32nd International
  Conference on Machine Learning (ICML-15)} (D.~Blei and F.~Bach, eds.),
  pp.~957--966, JMLR Workshop and Conference Proceedings, 2015.

\bibitem{sift}
D.~G. Lowe, ``{Distinctive Image Features from Scale-Invariant Keypoints},''
  {\em Int. J. Comput. Vision}, vol.~60, pp.~91--110, Nov. 2004.

\bibitem{tagspace}
J.~Weston, S.~Chopra, and K.~Adams, ``{\#}tagspace: Semantic embeddings from
  hashtags,'' in {\em Proceedings of the 2014 Conference on Empirical Methods
  in Natural Language Processing, {EMNLP} 2014, October 25-29, 2014, Doha,
  Qatar, {A} meeting of SIGDAT, a Special Interest Group of the {ACL}},
  pp.~1822--1827, 2014.

\bibitem{flann}
M.~Muja and D.~G. Lowe, ``Fast approximate nearest neighbors with automatic
  algorithm configuration,'' in {\em International Conference on Computer
  Vision Theory and Application VISSAPP'09)}, pp.~331--340, INSTICC Press,
  2009.

\bibitem{lsh}
M.~Datar, N.~Immorlica, P.~Indyk, and V.~S. Mirrokni, ``Locality-sensitive
  hashing scheme based on p-stable distributions,'' in {\em Proceedings of the
  Twentieth Annual Symposium on Computational Geometry}, SCG '04, (New York,
  NY, USA), pp.~253--262, ACM, 2004.

\bibitem{gong2011ITQ}
Y.~Gong and S.~Lazebnik, ``{Iterative Quantization: A Procrustean Approach to
  Learning Binary Codes},'' in {\em CVPR '11}.

\bibitem{lazebnik09}
M.~Raginsky and S.~Lazebnik, ``{Locality-sensitive binary codes from
  shift-invariant kernels},'' in {\em Advances in Neural Information Processing
  Systems 22} (Y.~Bengio, D.~Schuurmans, J.~Lafferty, C.~Williams, and
  A.~Culotta, eds.), pp.~1509--1517, Curran Associates, Inc., 2009.

\bibitem{wang12}
J.~Wang, S.~Kumar, and S.-F. Chang, ``{Semi-Supervised Hashing for Large-Scale
  Search},'' {\em IEEE Trans. Pattern Anal. Mach. Intell.}, vol.~34,
  pp.~2393--2406, Dec. 2012.

\bibitem{lin16}
K.~Lin, J.~Lu, C.-S. Chen, and J.~Zhou, ``Learning compact binary descriptors
  with unsupervised deep neural networks,'' in {\em CVPR}, 2016.

\bibitem{MAP}
P.~Dlugosch, D.~Brown, P.~Glendenning, M.~Leventhal, and H.~Noyes, ``An
  efficient and scalable semiconductor architecture for parallel automata
  processing,'' {\em Parallel and Distributed Systems, IEEE Transactions on},
  vol.~25, no.~12, pp.~3088--3098, 2014.

\bibitem{ap-hamming-distance}
Micron, ``{Calculating Hamming Distance}.''
  \url{http://www.micronautomata.com/documentation/cookbook/c_hamming_distance.html}.
\newblock Accessed: 2016-03-18.

\bibitem{race-logic}
A.~Madhavan, T.~Sherwood, and D.~Strukov, ``Race logic: A hardware acceleration
  for dynamic programming algorithms,'' in {\em Proceeding of the 41st Annual
  International Symposium on Computer Architecuture}, ISCA '14, (Piscataway,
  NJ, USA), pp.~517--528, IEEE Press, 2014.

\bibitem{spaghetti-sort}
A.~K. Dewdney, ``Computer recreations: On the spaghetti sort computer and other
  analog gadgets for problem solving,'' {\em Scientific American}, vol.~250(6),
  pp.~19--26, June 1984.

\bibitem{ap-personal}
M.~Grimm, M.~Tanner, and I.~Roy, ``Discussion with micron automata processor
  team.'' Personal Communication.

\bibitem{papadimitriou11}
K.~Papadimitriou, A.~Dollas, and S.~Hauck, ``Performance of partial
  reconfiguration in fpga systems: A survey and a cost model,'' {\em ACM Trans.
  Reconfigurable Technol. Syst.}, vol.~4, pp.~36:1--36:24, Dec. 2011.

\bibitem{alexnet}
A.~Krizhevsky, I.~Sutskever, and G.~E. Hinton, ``{ImageNet Classification with
  Deep Convolutional Neural Networks},'' in {\em Advances in Neural Information
  Processing Systems 25} (F.~Pereira, C.~Burges, L.~Bottou, and K.~Weinberger,
  eds.), pp.~1097--1105, Curran Associates, Inc., 2012.

\bibitem{pca}
I.~Jolliffe, ``Principal component analysis,'' 2014.

\bibitem{motion-planning}
A.~Yershova and S.~LaValle, ``{Improving Motion-Planning Algorithms by
  Efficient Nearest-Neighbor Searching},'' {\em Robotics, IEEE Transactions
  on}, vol.~23, pp.~151--157, Feb 2007.

\bibitem{how-many-results}
D.~Kelly and L.~Azzopardi, ``How many results per page?: A study of serp size,
  search behavior and user experience,'' in {\em Proceedings of the 38th
  International ACM SIGIR Conference on Research and Development in Information
  Retrieval}, SIGIR '15, (New York, NY, USA), pp.~183--192, ACM, 2015.

\bibitem{ap-technode}
K.~S. Mircea R.~Stan, ``Automata computing.'' University Lecture, 2015.

\bibitem{gpu-knn}
V.~Garcia, E.~Debreuve, and M.~Barlaud, ``{Fast k nearest neighbor search using
  GPU},'' in {\em Computer Vision and Pattern Recognition Workshops, 2008.
  CVPRW '08. IEEE Computer Society Conference on}, pp.~1--6, June 2008.

\bibitem{lshbox}
RSIA-LIESMARS-WHU, ``A c++toolbox for locality-sensitive hashing for large
  scale image retrieval.'' https://github.com/RSIA-LIESMARS-WHU/LSHBOX, 2015.

\bibitem{sparse-distributed-memory}
P.~Kanerva, {\em {Sparse Distributed Memory}}.
\newblock Cambridge, MA, USA: MIT Press, 1988.

\bibitem{pcam}
J.~D. Roberts, ``{PROXIMITY CONTENT-ADDRESSABLE MEMORY:AN EFFICIENT EXTENSION
  TO k-NEAREST NEIGHBORS SEARCH (M.S. Thesis)},'' tech. rep., Santa Cruz, CA,
  USA, 1990.

\bibitem{smart-access-memory}
A.~Lipman and W.~Yang, ``{The Smart Access Memory: An Intelligent RAM for
  Nearest Neighbor Database Searching},'' in {\em In ISCA Workshop on Mixing
  Logic and DRAM}, 1997.

\bibitem{tandon13}
P.~Tandon, J.~Chang, R.~G. Dreslinski, V.~Qazvinian, P.~Ranganathan, and T.~F.
  Wenisch, ``Hardware acceleration for similarity measurement in natural
  language processing,'' in {\em Proceedings of the 2013 International
  Symposium on Low Power Electronics and Design}, ISLPED '13, (Piscataway, NJ,
  USA), pp.~409--414, IEEE Press, 2013.

\bibitem{shinde10}
R.~Shinde, A.~Goel, P.~Gupta, and D.~Dutta, ``Similarity search and locality
  sensitive hashing using ternary content addressable memories,'' in {\em
  Proceedings of the 2010 ACM SIGMOD International Conference on Management of
  Data}, SIGMOD '10, (New York, NY, USA), pp.~375--386, ACM, 2010.

\bibitem{bremler-barr15}
A.~Bremler-Barr, Y.~Harchol, D.~Hay, and Y.~Hel-Or, ``Ultra-fast similarity
  search using ternary content addressable memory,'' in {\em Proceedings of the
  11th International Workshop on Data Management on New Hardware}, DaMoN'15,
  (New York, NY, USA), pp.~12:1--12:10, ACM, 2015.

\bibitem{ap-random-forest}
T.~T. II, Y.~Fu, I.~Roy, E.~Jonas, and P.~Glendenning, ``Towards machine
  learning on the automata processor,'' in {\em High Performance Computing -
  31st International Conference, {ISC} High Performance 2016, Frankfurt,
  Germany, June 19-23, 2016, Proceedings}, pp.~200--218, 2016.

\bibitem{ly16}
T.~Ly, R.~Sarkar, K.~Skadron, and S.~T. Acton, ``Feature extraction and image
  retrieval on an automata structure,'' in {\em Proceedings of the 50th
  Asilomar Conferences on Signals, Systems, and Computers}, Nov 2016.

\bibitem{zhang14}
D.~Zhang, N.~Jayasena, A.~Lyashevsky, J.~L. Greathouse, L.~Xu, and
  M.~Ignatowski, ``Top-pim: Throughput-oriented programmable processing in
  memory,'' in {\em Proceedings of the 23rd International Symposium on
  High-performance Parallel and Distributed Computing}, HPDC '14, (New York,
  NY, USA), pp.~85--98, ACM, 2014.

\bibitem{hsieh16}
K.~Hsieh, E.~Ebrahim, G.~Kim, N.~Chatterjee, M.~O'Connor, N.~Vijaykumar,
  O.~Mutlu, and S.~W. Keckler, ``Transparent offloading and mapping (tom):
  Enabling programmer-transparent near-data processing in gpu systems,'' in
  {\em 2016 ACM/IEEE 43rd Annual International Symposium on Computer
  Architecture (ISCA)}, pp.~204--216, June 2016.

\bibitem{farmahini-farahani15}
A.~Farmahini-Farahani, J.~H. Ahn, K.~Morrow, and N.~S. Kim, ``Nda: Near-dram
  acceleration architecture leveraging commodity dram devices and standard
  memory modules,'' in {\em 2015 IEEE 21st International Symposium on High
  Performance Computer Architecture (HPCA)}, pp.~283--295, Feb 2015.

\bibitem{ndp-graph}
J.~Ahn, S.~Hong, S.~Yoo, O.~Mutlu, and K.~Choi, ``A scalable
  processing-in-memory accelerator for parallel graph processing,'' in {\em
  Proceedings of the 42Nd Annual International Symposium on Computer
  Architecture}, ISCA '15, (New York, NY, USA), pp.~105--117, ACM, 2015.

\bibitem{pim-graphics}
C.~Xie, S.~L. Song, J.~Wang, W.~Zhang, and X.~Fu, ``Processing-in-memory
  enabled graphics processors for 3d rendering,'' in {\em 2017 IEEE 23rd
  International Symposium on High Performance Computer Architecture (HPCA)},
  Feb 2017.

\end{thebibliography}

\end{document}